\documentclass[12pt]{article}%
\usepackage{amsmath,latexsym}
\usepackage{graphicx}
\usepackage{amsmath}
\usepackage{amsfonts}
\usepackage{amssymb}%
\begin{document}

\title{{Revisiting the boundary conditions for a 
   Morris-Thorne wormhole}}
   \author
{Peter K.F. Kuhfittig*\\  \footnote{kuhfitti@msoe.edu}
 \small Department of Mathematics, Milwaukee School of
Engineering,\\
\small Milwaukee, Wisconsin 53202-3109, USA}

\date{}
 \maketitle
 
\begin{abstract}\noindent
In physical science, the concept of 
\emph{emergence} is often used to describe 
phenomena that occur at macroscopic scales 
but not at microscopic scales.  The latter 
is usually referred to as a \emph{fundamental 
property} and the former as an \emph{emergent
property}.  In this paper, noncommutative 
geometry, often viewed as an offshoot of 
string theory, is the primary fundamental 
theory that gives rise to macroscopic 
wormholes and their properties, thereby 
becoming an emergent phenomenon.  As a 
consequence of these considerations, we 
will reexamine the boundary conditions that 
characterize a Morris-Thorne wormhole.  The 
result is a significant modification of the 
wormhole structure.  
\end{abstract}

\section{Introduction}\label{E:introduction} 
In a previous paper \cite{pK22}, the author 
argued that a Morris-Thorne wormhole is 
necessarily a compact stellar object and, 
coupled with the concomitant relativistic 
effects, could sustain a sufficiently large 
wormhole without relying on exotic matter, 
a requirement that many researchers consider 
to be unphysical.  The purpose of this paper 
is to strengthen these conclusions by revisiting 
the boundary conditions for a Morris-Thorne 
wormhole, thereby calling for a modification 
of the wormhole structure: if $r=r_0$ is the 
throat, then the interior region $r<r_0$, while 
not part of the wormhole spacetime, can still 
have a significant effect caused by the enormous 
increase in the mass due to the relativistic 
effects stemming from its central location.  
The high radial tension is a direct consequence 
thereof.  The qualitative results are confirmed 
by invoking a noncommutative-geometry background.

As indicated in the Abstract, the 
noncommutative-geometry background is viewed 
as a fundamental property. The macroscopic 
wormhole and its properties are thereby 
emergent.

\section{Emergence}\label{S:emergence}
The concept of emergence is all around us. 
Whether we are talking about ant colonies 
or human consciousness, emergence describes 
phenomena that occur at macroscopic scales 
but not at microscopic scales.  In spite of 
these complexities, emergence has a certain 
intuitive appeal that seems to have had its 
origins in antiquity.  The basic idea can 
best be summarized by saying that the 
growing complexity causes the appearance 
of new features that are often unexpected 
and therefore surprising.  In other words, 
the new features do not appear to follow 
from more fundamental properties and emerge 
only with increasing interactions.  For 
example, life emerges from lifeless objects 
such as atoms and molecules.  This process 
is not reversible, however: living organisms 
do not tell us anything about the particles 
in the fundamental theory.  So by definition, 
\emph{emergent phenomena} are derived from 
some \emph{fundamental theory}.  As another 
example, the complex structure of an ant 
colony cannot be explained by the behavior 
of individual ants: the colony is an emergent 
phenomenon. 
 
Of particular interest to us are the physical 
properties that occur on macroscopic scales 
but not on microscopic scales, even though 
a macroscopic system consists of a large 
collection of microscopic systems; the 
emergent macroscopic theory illustrates 
the characteristic irreversibility.  So for 
our purposes, quantum field theory is the 
fundamental theory.  Some of these ideas 
will be discussed further in the next 
section.     

\section{Noncommutative geometry}
      \label{S:noncommutative}
Noncommutative geometry, an offshoot of 
string theory, is a viable approach to 
quantum gravity.  Here we assume  that 
point-like particles are replaced by
smeared objects, an assumption that 
is consistent with the Heisenberg 
uncertainty principle.  This approach
also helps to eliminate the divergences  
that normally occur in general relativity 
\cite{SS03, NSS06, NS10}.  It is shown in 
Ref. \cite{NSS06} that this goal can be met 
by assuming that spacetime can be encoded
in the commutator $[\textbf{x}^{\mu},\textbf{x}^{\nu}]
=i\theta^{\mu\nu}$, where $\theta^{\mu\nu}$ is
an antisymmetric matrix that determines the
fundamental cell discretization of spacetime
in the same way that Planck's constant $\hbar$
discretizes phase space.  According to Refs. 
\cite{NM08, LL12}, the smearing can be modeled
by using a so-called Lorentzian distribution 
of minimal length $\sqrt{\beta}$ instead of 
the Dirac delta function: the energy density 
$\rho$ of a static and spherically symmetric
and particle-like gravitational source has 
the form  
\begin{equation}\label{E:density}
   \rho(r)=\frac{m\sqrt{\beta}}{\pi^2
   (r^2+\beta)^2}.
\end{equation} 
According to this model, the gravitational 
source causes the mass $m$ to be diffused 
throughout the region of linear dimension 
$\sqrt{\beta}$ due to the uncertainty.

This behavior suggests that noncommutative 
geometry is a good candidate for the
fundamental theory briefly mentioned at 
the end of Sec. \ref{S:emergence}.

\section{Traversable wormholes}
   \label{S:traversable}
Wormholes are handles or tunnels in 
spacetime connecting widely separated 
regions of our Universe or entirely 
different universes.  Wormholes are as 
good a prediction of Einstein's theory 
as black holes, but they are subject to 
severe restrictions from quantum field 
theory.  In particular, holding a wormhole 
open requires a violation of the null 
energy condition, calling for the 
existence of ``exotic matter" \cite{MT88},
a requirement that many researchers 
consider to be completely unphysical.  The 
author has argued in Ref. \cite{pK22} that 
a wormhole is a compact stellar object 
whose relativistic effects could remove 
the need for exotic matter provided that 
the throat radius is sufficiently large, 
an additional condition that is eliminated 
in the present paper.

The line element for a Morris-Thorne 
wormhole is given by
\begin{equation}\label{E:line1}
  ds^{2}=-e^{2\Phi(r)}dt^{2}+\frac{dr^2}
  {1-\frac{b(r)}{r}}
  +r^{2}(d\theta^{2}+\text{sin}^{2}\theta\,
  d\phi^{2}),
\end{equation}
using units in which $c=G=1$ \cite{MT88}.
The motivation for this line element comes 
from Ref. \cite{MTW}:
\begin{equation*}
  ds^{2}=-e^{2\Phi(r)}dt^{2}+\frac{dr^2}
  {1-\frac{2m(r)}{r}}
  +r^{2}(d\theta^{2}+\text{sin}^{2}\theta\,
  d\phi^{2}),\quad r\le R
  \end{equation*}
  \begin{equation}\label{E:line2}
  =-\left(1-\frac{2M}{r}\right)dt^2
  +\frac{dr^2}{1-\frac{2M}{r}}
  +r^{2}(d\theta^{2}+\text{sin}^{2}\theta\,
  d\phi^{2}), \quad r>R.
\end{equation}
Here $m(r)$ is the effective mass inside
radius $r$, while $M$ is the mass of a star of
radius $R$ as seen by a distant observer.
If $\rho(r)$ is the energy density, then
the total mass-energy inside radius $r$
is given by
\begin{equation}
   m(r)=\int^r_04\pi (r')^2\rho(r')\,dr',
   \quad m(0)=0.
\end{equation}
In line element (\ref{E:line1}), $\Phi=\Phi(r)$ 
is called the \emph{redshift function}, which 
must be finite everywhere to prevent the 
occurrence of an event horizon.  The function 
$b=b(r)$ is called the \emph{shape function} 
since it determines the spatial shape of the
wormhole when viewed, for example, in an
embedding diagram \cite{MT88}.  The spherical
surface $r=r_0$ is called the \emph{throat}
of the wormhole, where $b(r_0)=r_0$, one of 
the conditions to be discussed further below. 
Additional requirements are $b'(r_0)<1$, called 
the \emph{flare-out condition}, $b(r)<r$ for 
$r>r_0$, and $b'(r_0)>0$.  Another requirement 
is asymptotic flatness: $\text{lim}_{r\rightarrow
\infty}\Phi(r)=0$ and $\text{lim}_{r\rightarrow
\infty}b(r)/r=0$.

A critical issue discussed in this paper and 
in Ref. \cite{pK22} is the flare-out condition 
and its consequences: this condition can only 
be met by violating the null energy condition 
(NEC), which states that
\begin{equation}
  T_{\alpha\beta}k^{\alpha}k^{\beta}\ge 0
\end{equation}
for all null vectors $k^{\alpha}$, where
$T_{\alpha\beta}$ is the energy-momentum
tensor.  As noted above, matter that violates 
the NEC is called ``exotic" in Ref. \cite{MT88}.  
In particular, for the outgoing null vector
$(1,1,0,0)$, the violation reads
\begin{equation}\label{E:violation}
   T_{\alpha\beta}k^{\alpha}k^{\beta}=
   \rho +p_r<0.
\end{equation}
Here, $T^t_{\phantom{tt}t}=-\rho$ is the
energy density, $T^r_{\phantom{rr}r}= p_r$
is the radial pressure, and
$T^\theta_{\phantom{\theta\theta}\theta}=
T^\phi_{\phantom{\phi\phi}\phi}=p_t$ is
the lateral (transverse) pressure.  Next,
let us list the Einstein field equations:

\begin{equation}\label{E:Einstein1}
  \rho(r)=\frac{b'}{8\pi r^2},
\end{equation}
\begin{equation}\label{E:Einstein2}
   p_r(r)=\frac{1}{8\pi}\left[-\frac{b}{r^3}+
   2\left(1-\frac{b}{r}\right)\frac{\Phi'}{r}
   \right],
\end{equation}
and
\begin{equation}\label{E:Einstein3}
   p_t(r)=\frac{1}{8\pi}\left(1-\frac{b}{r}\right)
   \left[\Phi''-\frac{b'r-b}{2r(r-b)}\Phi'
   +(\Phi')^2+\frac{\Phi'}{r}-
   \frac{b'r-b}{2r^2(r-b)}\right].
\end{equation}

Before continuing, we need to emphasize the 
connection between the violation of the NEC 
and the flare-out condition at the throat:
observe that from Eqs. (\ref{E:violation}), 
(\ref{E:Einstein1}), and (\ref{E:Einstein2}), 
we deduce that 
\begin{equation}\label{E:exotic}
   8\pi[\rho(r_0)+p_r(r_0)]=\frac{b'(r_0)
   -b(r_0)/r_0}{r_0^2}<0,
\end{equation}
since $b(r_0)=r_0$.  Given that the radial
tension $\tau(r)$ is the negative of $p_r(r)$, 
Eq. (\ref{E:violation}) can be written as
\begin{equation}\label{E:tension}
   \tau-\rho c^2>0,
\end{equation}
temporarily reintroducing $c$.  This 
inequality is the reason for the designation
``exotic matter" since $\tau>\rho c^2$
implies that there is an enormous radial
tension at the throat.  For further discussion 
of this problem, see Refs.
\cite{pK13, pK20, pK21, pK22a, pK22b}, as well 
as Sec. \ref{S:non}.

\section{The energy density $\rho(r)$ and 
   the flare-out condition}

We can see from Eq. (\ref{E:Einstein1}) that
\begin{equation}
   b(r)=r_0+\int^r_{r_0}8\pi (r')^2\rho(r')\,dr',
\end{equation}
confirming the boundary condition $b(r_0)=r_0$, 
noted in Sec. \ref{S:traversable}.  It also 
follows from Eqs. (\ref{E:line1}) and 
(\ref{E:line2}) that
\begin{equation}\label{E:mass}
   b(r)=2 m(r).
\end{equation}
To study the flare-out condition, we need to 
recall that $\rho(r)$ is likely to be very 
small in geometrized units.  So we normally 
have
\begin{equation}
    b'(r_0)=8\pi r_0^2\rho(r_0)<1,
\end{equation}
as desired.  To show that the assumption 
regarding $\rho$ is realistic, suppose we 
try $\rho(r_0)=10^{-9}\text{m}^{-2}$.  Then
\[
   \rho(r_0)=10^{-9}\frac{c^2}{G}
   \approx10^{18}\frac{\text{kg}}{\,\text{m}^3},
\]
which corresponds to nuclear matter.  So 
$\rho$ could be even smaller than $10^{-9}
\,\text{m}^{-2}$.  The consequences will be 
taken up in Sec. \ref{S:boundary}.

\section{Wormholes as emergent phenomena}

Discussions of emergence often emphasize 
the surprising or unexpected nature of the 
outcome.  Our results are no exceptions.  

The term ``wormhole" was coined by John A. 
Wheeler in the 1950's and referred to 
microscopic wormholes as potential models 
of elementary charged particles, thereby 
suggesting the possibility of macroscopic 
wormholes as emergent phenomena.

Another example involves the NEC: we can 
see from Eq. (\ref{E:violation}) that
\begin{multline}\label{E:violation2}
   T_{\alpha\beta}k^{\alpha}k^{\beta}=\rho(r)+p_r(r)
   =\frac{m\sqrt{\beta}}{\pi^2(r^2+\beta)^2}+
   \frac{1}{8\pi}\left[-\frac{b}{r^3}+
   2\left(1-\frac{b}{r}\right)\frac{\Phi'}{r}
   \right]_{r=r_0}\\
   =\frac{m\sqrt{\beta}}{\pi^2(r_0^2+\beta)^2}
   -\frac{1}{8\pi}\frac{1}{r_0^2}<0
\end{multline}
since $\sqrt{\beta}\ll 1$.  So the violation 
can be attributed to the noncommutative-geometry 
background, rather than some hypothetical 
``exotic matter," \emph{at least locally}.  Given 
that the radial tension $\tau$ is the negative 
of the radial pressure $p_r(r)$, $\rho(r_0)+
p_r(r_0)<0$ becomes $\tau-\rho c^2>0$ locally.
The emergent macroscopic property will be 
confirmed in the next section.

Next, let us to return to Eq. (\ref{E:Einstein1}) 
to determine the shape function:
 \begin{multline}\label{E:shape}
   b(r)=r_0+\int^r_{r_0}8\pi(r')^2\rho(r')dr'\\
   =\frac{4m}{\pi}
  \left[\text{tan}^{-1}\frac{r}{\sqrt{\beta}}
  -\sqrt{\beta}\frac{r}{r^2+\beta}-
  \text{tan}^{-1}\frac{r_0}{\sqrt{\beta}}
  +\sqrt{\beta}\frac{r_0}{r_0^2
  +\beta}\right]+r_0\\
  =\frac{4m}{\pi}\frac{1}{r}
  \left[r\,\text{tan}^{-1}\frac{r}{\sqrt{\beta}}
  -\sqrt{\beta}\frac{r^2}{r^2+\beta}-
  r\,\text{tan}^{-1}\frac{r_0}{\sqrt{\beta}}
  +\sqrt{\beta}\frac{r_0r}{r_0^2
  +\beta}\right]+r_0.
\end{multline}
We can now follow Ref. \cite{pK23}, which 
unexpectedly shows that $B=b/\sqrt{\beta}$
has the properties of a shape function.  
The reason is that $B$ can be readily 
expressed as a function of $r/\sqrt{\beta}$:
\begin{multline}\label{E:shape1}
   \frac{1}{\sqrt{\beta}}
   \,b(r)=
   B\left(\frac{r}{\sqrt{\beta}}\right)=\\
   \frac{1}{\sqrt{\beta}}\frac{4m}{\pi}
   \frac{\sqrt{\beta}}{r}\left[\frac{r}{\sqrt{\beta}}
   \,\text{tan}^{-1}\frac{r}{\sqrt{\beta}}
   -\frac{\left(\frac{r}{\sqrt{\beta}}\right)^2}
   {\left(\frac{r}{\sqrt{\beta}}\right)^2+1}
  -\frac{r}{\sqrt{\beta}}\,
  \text{tan}^{-1}\frac{r_0}{\sqrt{\beta}}
  +\frac{r}{\sqrt{\beta}}
  \frac{\frac{r_0}{\sqrt{\beta}}}
  {\left(\frac{r_0}{\sqrt{\beta}}\right)^2+1}
  \right]+\frac{r_0}{\sqrt{\beta}}.
\end{multline}
 We now have
 \begin{equation}\label{E:throat}
   B\left(\frac{r_0}{\sqrt{\beta}}\right)
   =\frac{r_0}{\sqrt{\beta}},
\end{equation}
the analogue of $b(r_0)=r_0$.  It follows 
that the throat size is macroscopic, 
confirming that we are indeed dealing 
with an emergent property.

\section{The boundary condition $b(r_0)=r_0$}
   \label{S:boundary}
We start this section by recalling the 
structure of Visser's thin-shell wormhole 
from a Schwarzschild black hole \cite{PV95}. 
Such a wormhole is constructed by taking two
copies of a Schwarzschild spacetime and
removing from each the four-dimensional region
\begin{equation}
   \Omega= \{r\le a\,|\,a>2M\},
\end{equation}
where $a$ is a constant \cite{PV95}.  By
identifying the boundaries, i.e., by letting
\begin{equation}
   \partial\Omega= \{r=a\,|\,a>2M\},
\end{equation}
we obtain a manifold that is geodesically
complete.  The condition $a>2M$ ensures 
that the wormhole spacetime is outside 
the event horizon.  

We know from Eq. (\ref{E:mass}) that 
$\frac{1}{2}b(r)$ is the effective mass 
inside radius $r$.  Since $r=r_0$ is the 
thoat of the wormhole, it follows from 
the definition of throat that the 
interior $r<r_0$ is outside the wormhole 
spacetime, suggesting that a Morris-Thorne 
wormhole has something in common with a 
thin-shell wormhole: $\frac{1}{2}b(r_0)=
m(r_0)$ must be the mass of the interior 
$r<r_0$; it would therefore be subject to 
relativistic effects.  We must first 
observe, however, that the mass of the 
interior, $\frac{1}{2}b(r_0)=\frac{1}{2}r_0$ 
appears to be impossible.  For example, in 
geometrized units, the mass of the Earth is 
0.44 cm, which is very much less than the 
radius.  As in Ref. \cite{pK22}, we will 
rescue the condition $b(r_0)=r_0$ by taking 
into account certain relativistic effects, 
thereby altering the structure of a 
Morris-Thorne wormhole.

Since $m(r)$ has units of length, it follows
from line element (\ref{E:line1}) that     
the element of volume is given by the 
relativistic form
\begin{equation}
   dV(r)=4\pi r^2\frac{1}
   {\sqrt{1-\frac{2 m(r)}{r}}}\,dr.
\end{equation}
Recalling that $m(r)$ is the effective 
mass inside radius $r$, we get 
\begin{equation}
   \frac{2m(r)}{r}=2\cdot\frac{4}{3}
   \frac{\pi r^3}{r}\rho(r)=\frac{8}{3}
   \pi r^2\rho(r).
\end{equation}It follows that
\begin{equation}
   dV(r)=4\pi r^2\frac{1}
   {\sqrt{1-\frac{8}{3}\pi r^2\rho(r)}}\,dr
\end{equation}
and 
\begin{equation}
  V(r)=\int^r_04\pi (r')^2\frac{1}
   {\sqrt{1-\frac{8}{3}\pi (r')^2\rho(r')}}
   \,dr',  
\end{equation}
the total volume inside $r=r_0$.  The total 
mass $M(r_0)$ inside $r=r_0$ is therefore
given by
\begin{equation}
   M(r_0)=\frac{V(r_0)}{r_0^2}=\frac{1}{r_0^2}
   \int^{r_0}_0 4\pi r^2\frac{1}
   {\sqrt{1-\frac{8}{3}\pi r^2\rho(r)}}
   \,dr.  
\end{equation}
Knowing that $\rho(r)$ is very small in 
geometrized units, let us obtain an estimate 
of $M(r_0)$ by letting $\rho(r)\rightarrow 0$:
\begin{equation}
   M(r_0)=\frac{1}{r_0^2}\int^{r_0}_0 4\pi r^2dr
   =\frac{4}{3}\pi r_0.
\end{equation}

To draw our conclusion, we will consider a
specific example of a mass inside $r=r_0$:
\begin{equation}
   \frac{3}{8\pi}M(r_0)=\frac{3}{8\pi}
   \left(\frac{4}{3}\pi r_0\right)=
   \frac{1}{2}r_0.
\end{equation}
Since $\frac{1}{2}b(r_0)$ is the mass of 
the interior $r<r_0$, we can let
\begin{equation}
   \frac{1}{2}b(r_0)=\frac{3}{8\pi}M(r_0)=
   \frac{1}{2}r_0,
\end{equation}
which implies that $b(r_0)=r_0$, the desired 
boundary condition.  So our qualitative approach 
implies that the relativistic mass could be 
large enough to meet the condition $b(r_0)=r_0$ 
 without hypothesizing the need for exotic 
 matter. 

The conclusion depends on having a sufficiently 
large mass inside $r=r_0$ to yield the desired 
relativistic effects.  So qualitatively, we 
can even argue that since we are dealing with 
a compact stellar object, the large radial 
tension from Inequality (\ref{E:exotic}) is 
a natural consequence.  In other words, 
Inequality (\ref{E:exotic}) can be rewritten 
as
\begin{equation}\label{E:exotic2}
   8\pi\left[\rho(r_0)+p_r(r_0)\right]=
   \frac{b'(r_0)-\frac{M(r_0)}{\frac{1}{2}
   r_0}}{r_0^2}<0.
\end{equation}   
So $\rho(r_0)+p_r(r_0)<0$ and hence 
$\tau(r_0)-\rho(r_0)>0$ are due entirely 
to the relativistic effects.  We will 
confirm this inequality in the next section 
by invoking a noncommutative-geometry 
background. 

\emph{Remark:} Our idealized solution 
shows that the condition $M(r_0)=
\frac{1}{2}r_0$ can be met due to the 
relativistic effects, but this necessarily 
restricts the throat size $r=r_0$ since 
an excessively large $M(r_0)$  could 
lead to gravitational collapse.

\section{Returning to noncommutative 
     geometry}\label{S:non}

We continue our discussion of the 
relativistic effects by invoking a
noncommutative-geometry background, 
as noted after Eq. (\ref{E:exotic2}).

The small value of $\rho$ allows us 
to retain our previous conclusion: 
$M(r_0)=\frac{1}{2}r_0$ and hence 
$b(r_0)=r_0$, as we saw in the 
previous section.  Our main goal in 
this section is to confirm Inequality 
(\ref{E:tension}), $\tau-\rho c^2>0$.

It is noted in Ref. \cite{pK20} that 
the throat $r=r_0$ is a smeared 
surface since it is made up entirely 
of smeared particles.  The energy 
density $\rho_{s}$ of the surface is 
given by
\begin{equation}\label{E:rho2}
   \rho_s=\frac{\mu\sqrt{\beta}}{\pi^2
   [(r-r_0)^2+\beta]^2},
\end{equation}
where $\mu$ is the mass of the surface.  
If $r=r_0$, we return to Eq. (\ref{E:density}).  
Eq. (\ref{E:rho2}) can also be interpreted 
as the energy density of the spherical 
surface, yielding a smeared mass of the shell 
in the outward radial direction, the analogue 
of the smeared mass at the origin.  According 
to Ref. \cite{NSS06}, the relationship 
between the radial pressure and the energy 
density is $p_r=-\rho$.  This carries over 
to $p_r=-\rho_{s}$ in the outward radial 
direction.  Since the tension $\tau$ is the 
negative of $p_r$, the violation of the 
NEC, $p_r+\rho_{s}<0$, now becomes 
$\tau-\rho_{s}>0$.  The condition 
$p_r=-\rho_{s}$ implies that we are 
right on the edge of violating the NEC, 
i.e., $\tau-\rho_{s}=0$.  At $r=r_0$, 
Eq. (\ref{E:rho2}) gives 
\begin{equation}\label{E:rho3}
   \rho_{s}=\frac{\mu}{\pi^2}
      \frac{1}{\beta^{3/2}}.
\end{equation} 
However, since the throat is a smeared 
surface, we only have $r\approx r_0$.  
By Eq. (\ref{E:rho2}), $\rho_{s}$ is 
thereby reduced.  So instead of 
$\tau-\rho_{s}=0$, we actually have 
$\tau-\rho_{s}>0$, confirming 
Inequality (\ref{E:tension}), 
$\tau-\rho c^2>0$.  To check the 
plausibility of Eq. (\ref{E:rho3}), 
let us assume that $\mu$ has the rather 
minute value of $10^{-10}$ g.  According 
to Ref. \cite{pK20}, for a throat size 
of 10 m, $\tau\approx 5\times 10^{41}
  \text{dyn}/\text{cm}^2$.
So
\begin{equation}
  \tau=\rho_sc^2=\frac{\mu}{\pi^2}
  (\sqrt{\beta})^{-3}c^2=5\times 10^{41}
  \,\frac{\text{dyn}}{\text{cm}^2}.
\end{equation} 
Solving for $\sqrt{\beta}$, we find 
that the value of $\sqrt{\beta}=10^{-11}$ cm 
is sufficient.  Since $\sqrt{\beta}$ may 
be much smaller, we could accommodate 
even larger values of $\tau$.  We 
conclude that thanks to our 
noncommutative-geometry background, 
the condition $\tau-\rho c^2>0$ can 
be met without the need for exotic 
matter.

\section{Summary}

The purpose of this paper is to use 
the concept of emergence to show that 
noncommutative geometry, viewed as a
fundamental phenomenon, gives rise to 
macroscopic wormholes, collectively 
viewed as an emergent phenomenon.  These 
considerations call for a reexamination 
of the boundary conditions of 
Morris-Thorne wormholes, resulting in a 
modification of the wormhole structure:
if $r=r_0$ is the throat, then the 
interior region $r<r_0$, while not part 
of the wormhole spacetime, can still 
have a significant effect caused by the 
enormous increase in the mass $M(r_0)$ 
due to the relativistic effects stemming 
from its central location.   The result   
is $M(r_0)=\frac{1}{2}r_0$, thereby yielding 
the key boundary condition $b(r_0)=r_0$ 
that characterizes a Morris-Thorne wormhole.  
The equally problematical high radial 
tension is a direct consequence thereof. 
We confirm the conclusions by invoking 
our fundamental phenomenon, the 
noncommutative-geometry background.
\\
\\
\textbf{Data Availability Statement}
\\
No new data were generated in support of this manuscript.

\end{document}